\documentclass[a4paper,10pt,twoside]{cpc-hepnp}
\usepackage{multicol}
\usepackage{graphicx}
\usepackage{booktabs}
\usepackage{amssymb,bm,mathrsfs,amscd}
\usepackage[tbtags]{amsmath}
\usepackage{lastpage}
\usepackage{braket,slashed}
\usepackage{multirow}
\usepackage{subfigure}
\usepackage{CJK}

\begin{document}
\allowdisplaybreaks
\providecommand{\url}[1]{}
\bibliographystyle{hepnp}
\fancyhead[co]{\footnotesize Study on the mechanism of open-flavor strong decays}
\footnotetext[0]{Received \today}

\title{Study on the mechanism of open-flavor strong decays
      }

\author{
   LI Bao-Fei$^1$%
  \quad CHEN Xiao-Lin$^2$%
  \quad DENG Wei-Zhen$^{1;1)}$\email{dwz@pku.edu.cn}}
\maketitle

\address{%
  1~(Department of Physics and State Key
  Laboratory of Nuclear Physics and Technology, \\
  Peking University, Beijing 100871, China)\\
  2~(Department of Physics, Peking University, Beijing 100871, China)\\
}

\begin{abstract}
  The open-flavor strong decays are studied based on the interaction
  of potential quark model. The decay process is related to the
  $s$-channel contribution of the same scalar confinement and
  one-gluon-exchange(OGE) interaction in the quark model. After we
  adopt the prescription of massive gluons in time-like region from
  the lattice calculation, the approximation of four-fermion
  interaction is applied. The numerical calculation is performed to
  the meson decays in $u$, $d$, $s$ light flavor sector. The analysis
  of the $D/S$ ratios of $b_1\rightarrow \omega \pi$ and
  $a_1\rightarrow \rho \pi$ shows that the scalar interaction should be
  dominant in the open-flavor decays.
\end{abstract}



\begin{multicols}{2}

\section{Introduction}

Although QCD is considered a correct theory for strong
interactions, knowledge about the hadron structure in low energy
region is restricted due to the color confinement.  The potential
quark model is widely used to identify the conventional hadron states
in hadron physics. In contrary to its impressive success in hadron
spectra, especially when heavy quarks are involved, its interpretation
to the hadron strong decays is unsatisfied. Although there exist some
phenomenological models for the strong decays, the relationship
between these models and the potential quark model is somewhat
obscure.

One of the most popular models for open-flavor strong decays is
${}^3P_0$ model developed in the 1970s \cite{L1,L2}. This model has
successfully demonstrated its universal practical utility when applied
to a great number of particular decay channels
\cite{L3,L4,L14,L19,L20}. Later the flux-tube-breaking model was
proposed and ${}^3P_0$ model could be regarded as a limiting case of
this improved model \cite{L5}.  As early as 1978, Eichten
et~al. \cite{L6} developed the Cornell model by incorporating the
possibility of creation of a light-quark pair into the quark model
Hamiltonian.  However, in their model they considered the quark
interaction the time-component part of the vector interaction and
assumed that the interaction of the quark pair creation was the same
as the instantaneous interaction between two constituent quarks.  In
recent years an extended model including the scalar confining and
vector OGE interactions was studied by E.S.Ackleh et~al. \cite{L7}.  A
similiar scalar color-singlet confining interaction was dervied from a
relativistic string breaking kernel \cite{Simonov:2011cm}.

The instantaneous interaction in the above models always assumes Breit
approximation when dealing with gluon's momentum. For the potentials
in quark model, the energy of the exchanged gluon is negligible as
compared with the the masses of constituent quarks. Therefore the
transferred gluon momentum is space-like. Nevertheless, it is in all
the probability time-like if considering the creation of
quark-antiquark pair by the gluons.  Besides, based on the recent study
in lattice field theory \cite{L8,L9}, gluons are supposed to act as
 massive vector bosons in non-perturbative region with masses
evaluated about 600$\sim$1000MeV.  A non-vanishing gluon mass is also
needed in the phenomenological calculation of the diffractive scattering
\cite{L21} and radiative decays of the $J/\psi$ and $\Upsilon$
\cite{L22}.

In this paper, an alternative study of the open-flavor strong decays
is taken and inspected by the experimental decay widths.  Following
Ref.~\cite{L7}, the quark pair-creation interaction consists of a
scalar confining interaction and an OGE part. We will distinguish the
Breit approximation of the gluon's propagator in the time-like region
from that in the space-like region.  In the time-like non-perturbative
region, the massive gluon prescription is adopted according to
Refs.~\cite{L8,L9}.  In this way, the decay interaction will be
further simplified to the form of four-fermion interaction.

\section{The Decay Model}

To describe the creation of a light-quark pair in the quark model, a
plausible approach is to consider the field quantization of the quark
potential.  In the Cornell model, the quark potential is replaced by
an instantaneous interaction \cite{L6}
\begin{equation}
  H_I = \frac12 \int d^3 x d^3 y :\rho_a(\bm{x}) \frac34
  V(\bm{x}-\bm{y}) \rho_a(\bm{y}):.
\end{equation}
where
\begin{equation}
  \rho_a(\bm{x}) = \sum_{\text{flavors}} \psi^\dag(\bm{x})
  T_a \psi(\bm{x}).
\end{equation}
is the quark color-charge-density operator.  Here $\psi(x)$ denotes the
quark field with flavor and color indices suppressed, and $T_a$ stands for
the Gell-Mann matrices for $SU(3)$ generators.  Since the confinement
should be the Lorentz scalar, in Ref.~\cite{L7} the instantaneous
interaction is replaced by the combination of the scalar confinement
interaction and the vector OGE interaction.

We will start from the covariant nonlocal current-current action of
the quark interaction \cite{L10}:
\end{multicols}
\ruleup
\begin{equation}
  \label{eq1}
  A = -\frac12\int d^4 x d^4 y \bar{\psi}(x) \gamma_\mu
  T_a\psi(x)  G(x-y) \bar{\psi}(y) \gamma^\mu T_a\psi(y)
   -\frac12\int d^4 x d^4 y \bar{\psi}(x)
  T_a\psi(x) S(x-y) \bar{\psi}(y) T_a\psi(y).
\end{equation}
\ruledown \vspace{0.5cm}
\begin{multicols}{2}
The vector kernel $G(x-y)$ corresponds to the gluon propagator in
coordinate space which generates the OGE Coulomb potential
$-\frac{\alpha_s}{r}$ in the quark model.  In the momentum space
\begin{equation}
  G(q^2) = -\frac{4\pi\alpha_s}{q^2}.
\end{equation}
On the other hand, the scalar kernel $S(x-y)$ should generate the
linear confining potential $\frac{3}{4}b r$. Thus in the momentum
space
\begin{equation}
  S(q^2) = -\frac{6\pi b}{q^4}.
\end{equation}
The relevant coupling constants $\alpha_s$ and $b$ are the potential
parameters in the potential quark model.

The lattice calculation shows that the behavior of the gluon
propagator is quite different in the non-perturbative region.  In
Refs.~ \cite{L8, L9}, the transverse propagator is assumed to be:
\begin{equation}
  D(q^2) = \frac{Z(q^2)}{q^2-M^2(q^2)}.
\end{equation}
where $M(q^2)$ is the running gluon mass. Then the kernels $G$ and $S$
are modified to
\begin{align}
  G(q^2) =& -\frac{4\pi\alpha_s}{q^2-M^2(q^2)}. \\
  S(q^2) =& -\frac{6\pi b}{[q^2-M^2(q^2)]^2}.
\end{align}
The lattice simulations suggest $M(0)=600\sim 1000$ MeV which means
that gluon gets a non-vanishing mass $M_g$ in the non-perturbative
region $q\ll\Lambda_\text{QCD}$.  If the $q^2$ term in the gluon's
propagator is neglected in the quark-antiquark pair-creation process,
then the decay interaction is simplified to the four-fermion
interaction. The interaction Hamiltonian density for pair-creation
turns to be:
\begin{equation}
  \mathcal{H}_I(x) =  \mathcal{H}_s(x) + \mathcal{H}_v(x).
\end{equation}
where $\mathcal{H}_s(x)$ and $\mathcal{H}_v(x)$ represent the scalar
and vector interaction respectively:
\begin{align}
  \mathcal{H}_s(x)=& \frac{3\pi b}{M_g^4}\bar{\psi}(x)T_a
  \psi(x)\bar{\psi}(x)T^a \psi(x) .\label{eq4-1} \\
  \mathcal{H}_v(x)=& -\frac{2\pi \alpha_s}{M_g^2}\bar{\psi}(x)
  \gamma_\mu T_a
  \psi(x)\bar{\psi}(x)\gamma^\mu T^a \psi(x) .\label{eq4-2}
\end{align}

From the interaction in Eqs.~(\ref{eq4-1}) and (\ref{eq4-2}), we
can derive the formulae for decay rates within non-relativistic limit. As
meson states are normalized to $2E$ in our work,
\begin{equation}
  \braket{\bm{p}|\bm{p}'} = 2E \delta^3(\bm{p}-\bm{p}').
\end{equation}
the differential decay width in two-body decay process $A\rightarrow
B+C$ is expressed in terms of transition amplitude as:
\begin{equation}
  d\Gamma=\frac{\mathcal{S}|\mathcal{M}|^2}{2E_A}(2\pi)^7\delta^4(P_A-P_B-P_C)
  \frac{d^3\bm{P}_B}{2E_B}\frac{d^3\bm{P}_C}{2E_C}.
\end{equation}
where $\mathcal{S}$ is the symmetric factor
\begin{equation}
  \mathcal{S}= \frac{1}{1+\delta(B,C)}.
\end{equation}
The amplitude $\mathcal{M}$ is
related to the decay interaction through:
\begin{equation}
  \mathcal{M}=<BC|\mathcal{H}_I(0)|A>=\mathcal{M}_v+\mathcal{M}_s.
\end{equation}
where $\mathcal{M}_v$ and $\mathcal{M}_s$ are the amplitudes from
vector and scalar interaction respectively.

\begin{figure}
\begin{center}
\caption{\label{Fig-1}%
  Contributions to vector interaction.}
\subfigure[]{\includegraphics[scale=0.4]{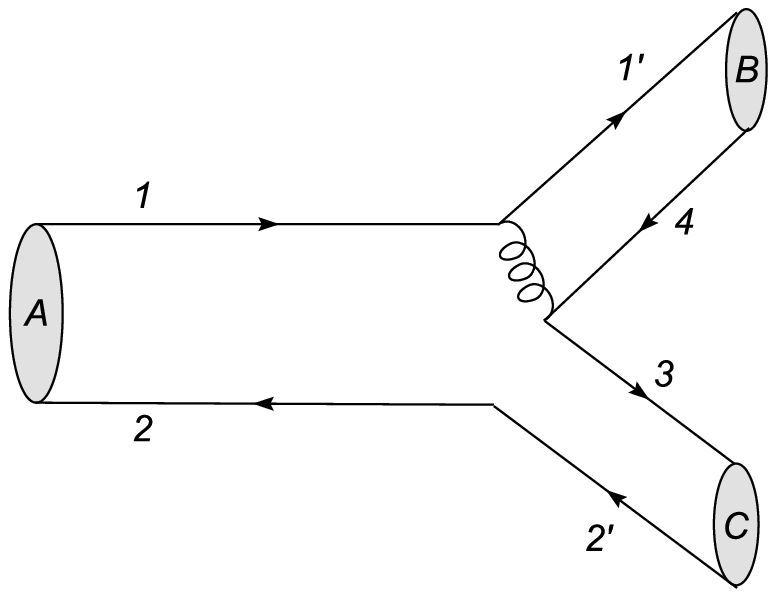}}\hspace{2cm}
\subfigure[]{\includegraphics[scale=0.4]{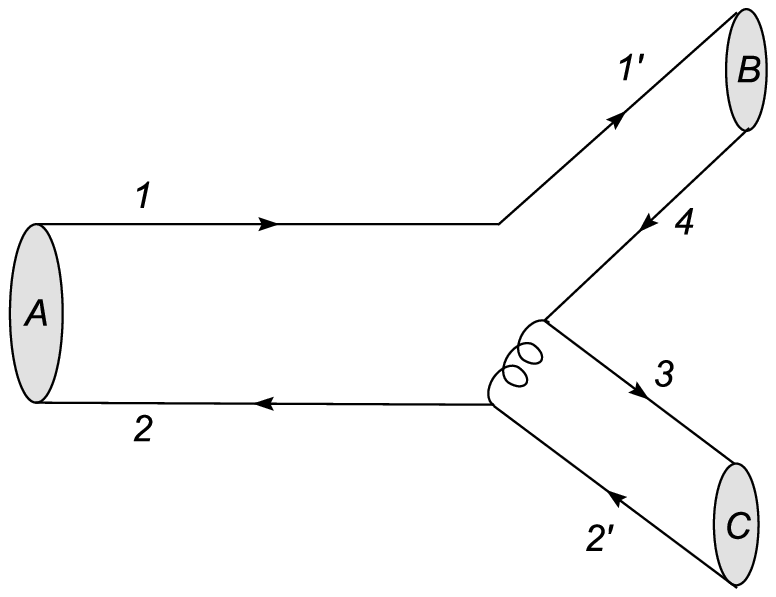}}\\
\subfigure[]{\includegraphics[scale=0.4]{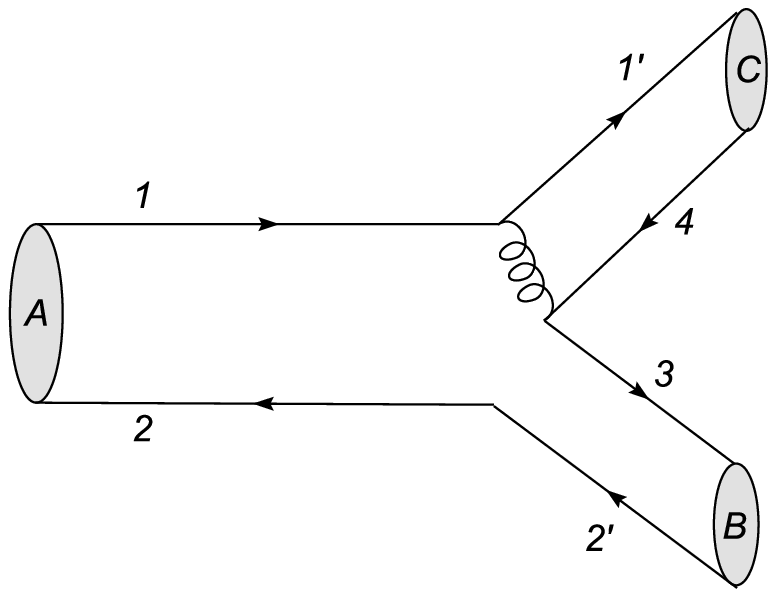}}\hspace{2cm}
\subfigure[]{\includegraphics[scale=0.4]{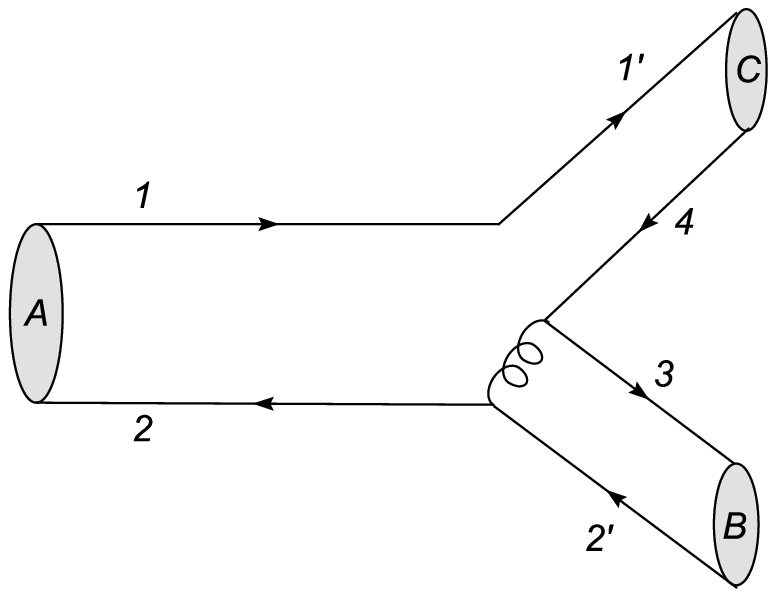}}
\end{center}
\end{figure}

For each interaction, the transition amplitude comes from four
diagrams. For the vector interaction, its Feynman diagrams are shown
in Figure~\ref{Fig-1}.
\end{multicols}
\begin{multicols}{2}
In Diagram (a) the $q\bar{q}$ pair is created from gluons emitted
by the initial quark while in Diagram (b) the $q\bar{q}$ pair is
created from gluons emitted by the initial anti-quark.  Diagrams
(c) and (d) come from the interchange of final particles B and C in
Diagrams (a) and (b) respectively.

The total decay width is expressed as
\begin{equation}
  \Gamma=\frac{16\pi^7\mathcal{S}P_f}{M_A^2}\sum_{L S}|\mathcal{M}^{LS}|^2.
\end{equation}
where $\bm{P}_f=\bm{P}_B=-\bm{P}_C$ in the rest frame of initial
particle $A$ and $\mathcal{M}^{LS}$ are the partial wave amplitudes.

For further simplification, the space wave functions of all meson
states are taken to be the simple harmonic oscillator (SHO) wave
functions with a common oscillator parameter $\beta$. The partial wave
amplitudes $\mathcal{M}^{LS}$ are presented in the appendix.

\section{Results and Analysis}

In the numerical calculation, the common oscillator parameter
$\beta=400$MeV is adopted from Ref.~\cite{L10}. All related
masses of mesons are taken from Ref.~\cite{L11}. Other parameters,
like constituent quark masses, coupling constants $\alpha_s$ and $b$
are also taken from Ref.~\cite{L10}. They are $m_u=m_d=220$MeV,
$m_s=419$MeV, $\alpha_s=0.60$, and $b=0.18\text{GeV}^2$
respectively.

The only one parameter which cannot be determined from the quark model
is the effective gluon mass $M_g$. In this work, this parameter is
fixed in a least square fit to the experimental decay widths. We find
that $M_g=668$MeV which falls within the range $600\sim1000$MeV
estimated in the lattice calculation~\cite{L8,L9}.

The results of the decay widths are tabulated in Table~\ref{Table-2}
together with the decay widths of ${}^3P_0$ model and the experimental
data.  In the table, $\Gamma_1$ indicates the decay rates of our
calculation. As can been seen, the widths of the decay processes
characterized by the creation of $s\bar s$ pair are rather small
compared with the experimental data. The reason is that the creation of
$s\bar s$ pair is suppressed in the four-fermion interaction due to
the heavier mass of $s$ quark. Note that in the ${}^3P_0$ model, the
transition operator is independent of the flavor mass in $q\bar q$
pair creation. This shows that the $q\bar q$ pair creation has the
flavor $SU(3)$ symmetry. In the third column $\Gamma_2$ in
Table~\ref{Table-2}, the decay widths related to $s\bar s$ pair
creation are recalculated with the flavor symmetry restored with
$m_s=m_u=220$MeV in the decay Hamiltonian. The corresponding decay
widths are enhanced which improves the fit to the experimental data.
\end{multicols}
\ruleup
\begin{center}
\tabcaption{\label{Table-2}The decay widths.
    The decay widths of ${}^3P_0$ model are taken from Ref.~\cite{L3}.
    The experimental data are taken from Ref.~\cite{L11}. Unit: MeV}
\footnotesize
\begin{tabular*}{170mm}{@{\extracolsep{\fill}}cccccc}
     \toprule Channel&$\Gamma_1$&$\Gamma_2$&$\Gamma_3$&${}^3P_0$&Exp\\\hline
      $\rho\rightarrow\pi\pi$&109.7&&138&96&149\\
      $b_1\rightarrow\omega\pi$&57.7&&160&176&142\\\
      $a_2\rightarrow\rho\pi$&51.5&&49.7&65&75.4\\
      $a_2\rightarrow\eta\pi$&15.4&&19.5&&15.5\\
      $a_2\rightarrow K\bar{K}$&1.64&9.03&2.26&11&5.24\\
      $a_2\rightarrow\eta'\pi$&1.20&&1.52&&0.567\\
      $\pi_2\rightarrow f_2\pi$&58.6&&77.1&147&146\\
      $\pi_2\rightarrow\rho\pi$&58.3&&128&232&80.3\\
      $\pi_2\rightarrow K^*\bar{K}+c.c.$&0.23&10.8&4.66&38&10.9\\
      $\pi_2\rightarrow\rho\omega$&9.23&&25.1&&7.00\\
      $\rho_3\rightarrow\pi\pi$&47.3&&59.6&116&38.0\\
      $\rho_3\rightarrow\omega\pi$&17.8&&17.2&36&25.8\\
      $\rho_3\rightarrow K\bar{K}$&0.68&7.06&0.94&9.2&2.54\\
      $f_2\rightarrow \pi\pi$&136&&172&203&157\\
      $f_2\rightarrow K\bar{K}$&1.07&5.98&1.48&7.2&8.51\\
      $f_4\rightarrow \omega\omega$&23.9&&14.9&53&54\\
      $f_4\rightarrow \pi\pi$&27.6&&34.8&123&40.3\\
      $f_4\rightarrow K\bar K$&0.18&3.45&0.25&5.4&1.61\\
      $f_0(1500)\rightarrow \pi\pi$&108&&34.7&&38.0\\
      $f_0(1500)\rightarrow K\bar{K}$&4.95&7.99&0.49&&9.38\\
      $\phi\rightarrow K^+K^-$&1.96&2.63&2.18&2.37&2.10\\
      $f'_2\rightarrow K\bar{K}$&84.3&78.9&21.0&117&64.8\\
      $K^*\rightarrow K\pi$&41.7&45.9&46.4&36&50.8\\
      $K^*(1410)\rightarrow K\pi$&21.7&32.4&1.15&&15.3\\
      $K^*_0\rightarrow K\pi$&348&1062&194&163&251\\
      $K^*_2\rightarrow K\pi$&81.7&73&90.2&108&49.2\\
      $K^*_2\rightarrow K^*\pi$&23.0&20.3&20.6&27&24.3\\
      $K^*_2\rightarrow K\rho$&7.22&6.20&6.42&9.3&8.57\\
      $K^*_2\rightarrow K\omega$&2.12&1.82&1.88&2.6&2.86\\
      $K^*_3\rightarrow K\rho$&14.4&9.54&12.5&24&49.3\\
      $K^*_3\rightarrow K^*\pi$&19.5&13.8&17.7&33&31.8\\
      $K^*_3\rightarrow K\pi$&45.1&32&49.4&87&30.0\\
      $K^*_4\rightarrow K\pi$&18.4&10.1&20&55&19.6\\
      $K^*_4\rightarrow K^*\phi$&0.37&2.44&0.23&3.2&2.8\\
    \bottomrule
    \end{tabular*}
\vspace{6mm}
\ruledown
\end{center}
\begin{multicols}{2}
Individual decay amplitudes from scalar and vector interactions are
listed in Table~\ref{Table-3}.  The scalar interaction is dominant in
most of the decay channels.  However in the channels $1D\rightarrow
1P+1S$, $2S\rightarrow 1S+1S$ and $2P\rightarrow 1S+1S$ the
contribution from the vector interaction is important, while in the
channel ${}^3P_0\rightarrow{}^1S_0+{}^1S_0$ the vector interaction
becomes dominant, as can be seen in the process $f_0(1370)\rightarrow
\pi\pi$ whose decay width may amount to $1000$ MeV which is too large
compared with the experimental result of $200-500$MeV.
\end{multicols}
\ruleup
\begin{center}
\tabcaption{\label{Table-3}The individual amplitudes from scalar and
    vector interactions. Unit: MeV}
\footnotesize
\begin{tabular*}{170mm}{@{\extracolsep{\fill}}cccccc}
     \toprule  channel&L&S&$M_s$&$M_v$&M\\
      \hline
      $\rho\rightarrow\pi\pi$&P&0&-1.39&-0.55&-1.94\\
      $b_1\rightarrow\omega\pi$&S&1&2.34&-0.35&2.00\\
      $b_1\rightarrow\omega\pi$&D&1&0.62&0.49&1.11\\
      $a_2\rightarrow\rho\pi$&D&1&-1.32&-0.79&-2.11\\
      $a_2\rightarrow\eta\pi$&D&0&-0.73&-0.29&-1.02\\
      $a_2\rightarrow K\bar{K}$&D&0&-0.28&-0.09&-0.36\\
      $a_2\rightarrow\eta'\pi$&D&0&-0.28&-0.11&-0.39\\
      $\pi_2\rightarrow f_2\pi$&S&2&2.32&-5.52&-3.20\\
      $\pi_2\rightarrow f_2\pi$&D&2&0.383&-0.003&0.380\\
      $\pi_2\rightarrow f_2\pi$&G&2&0.01&0.004&0.014\\
      $\pi_2\rightarrow\rho\pi$&P&1&1.85&-0.72&1.14\\
      $\pi_2\rightarrow\rho\pi$&F&1&1.10&0.88&1.98\\
      $\pi_2\rightarrow K^*\bar{K}+c.c.$&P&1&0.49&-0.56&-0.07\\
      $\pi_2\rightarrow K^*\bar{K}+c.c.$&F&1&0.09&0.07&0.16\\
      $\pi_2\rightarrow\rho\omega$&P&1&1.39&-0.09&1.31\\
      $\pi_2\rightarrow\rho\omega$&F&1&0.13&0.10&0.23\\
      $\rho_3\rightarrow\pi\pi$&F&0&-1.31&-0.52&-1.83\\
      $\rho_3\rightarrow\omega\pi$&F&1&-0.80&-0.47&-1.27\\
      $\rho_3\rightarrow K\bar{K}$&F&0&-0.18&-0.06&-0.24\\
      $f_2\rightarrow \pi\pi$&D&0&-1.95&-0.77&-2.72\\
      $f_2\rightarrow K\bar{K}$&D&0&-0.23&-0.08&-0.31\\
      $f_4\rightarrow \omega\omega$&G&0&-0.08&-0.03&-0.11\\
      $f_4\rightarrow \omega\omega$&D&2&0.88&0.87&1.75\\
      $f_4\rightarrow \omega\omega$&G&2&0.15&0.15&1.30\\
      $f_4\rightarrow \pi\pi$&G&0&-1.09&-0.43&-1.52\\
      $f_4\rightarrow K\bar K$&G&0&-0.10&-0.03&-0.13\\
      $f_0(1500)\rightarrow \pi\pi$&S&0&0.95&-3.56&-2.61\\
      $f_0(1500)\rightarrow K\bar{K}$&S&0&0.13&-0.77&-0.64\\
      $\phi\rightarrow K^+ K^-$&P&0&-0.40&-0.20&-0.60\\
      $f'_2\rightarrow K\bar{K}$&D&0&-0.84&-0.28&-1.12\\
      $K^*\rightarrow K\pi$&P&0&-1.04&-0.50&-1.54\\
      $K^*(1410)\rightarrow K\pi$&P&0&-0.18&1.39&1.21\\
      $K^*_0\rightarrow K\pi$&S&0&-2.32&7.19&4.87\\
      $K^*_2\rightarrow K\pi$&D&0&-1.58&-0.78&-2.36\\
      $K^*_2\rightarrow K^*\pi$&D&1&0.92&0.61&1.52\\
      $K^*_2\rightarrow K\rho$&D&1&-0.59&-0.39&-0.98\\
      $K^*_2\rightarrow K\omega$&D&1&-0.32&-0.21&-0.53\\
      $K^*_3\rightarrow K\rho$&F&1&-0.74&-0.50&-1.24\\
      $K^*_3\rightarrow K^*\pi$&F&1&0.74&0.44&1.18\\
      $K^*_3\rightarrow K\pi$&F&0&-1.27&-0.63&-1.90\\
      $K^*_4\rightarrow K\pi$&G&0&-0.86&-0.43&-1.29\\
      $K^*_4\rightarrow K^*\phi$&G&0&-0.003&-0.002&-0.005\\
      $K^*_4\rightarrow K^*\phi$&D&2&0.15&0.15&0.30\\
      $K^*_4\rightarrow K^*\phi$&G&2&0.007&0.007&0.014\\
      \bottomrule
    \end{tabular*}
\vspace{6mm}
\ruledown
\end{center}
\begin{multicols}{2}
One of the important criteria for the strong decay models is the $D/S$
amplitude ratios in the decays $b_1\rightarrow \omega\pi$ and
$a_1\rightarrow \rho\pi$. Experimentally, these ratios are
$D_1=0.277\pm0.027$ and $D_2=-0.062\pm0.02$ respectively
\cite{L11}. In the current model, analytic expressions for these
ratios are:
\begin{eqnarray}
  D_1&=&-\frac{\sqrt{2}p^2_f(15b+8m_g^2\alpha)}
  {15bp^2_f+8m^2_gp^2_f\alpha-72b\beta^2} .\\
  D_2&=&\frac{\sqrt{2}p^2_f(5b+2\alpha m^2_g)}
  {2b(5p^2_f-24\beta^2)+4\alpha m^2_g(p^2_f+8\beta^2)}  .\label {eq35}
\end{eqnarray}

According to the preceding values of the model parameters, the ratios'
numerical results are: $D_1=0.566$ and $D_2=0.731$. It is apparent
that $D_1$ is about two times larger than the experimental value while
the calculated value of $D_2$ has a wrong sign. Since the $\beta$
value is dependent on the wave function of meson, in
Figure~\ref{Figure-1} we show the dependence of the ratios on $\beta$.
\begin{center}
 \includegraphics[scale=0.5]{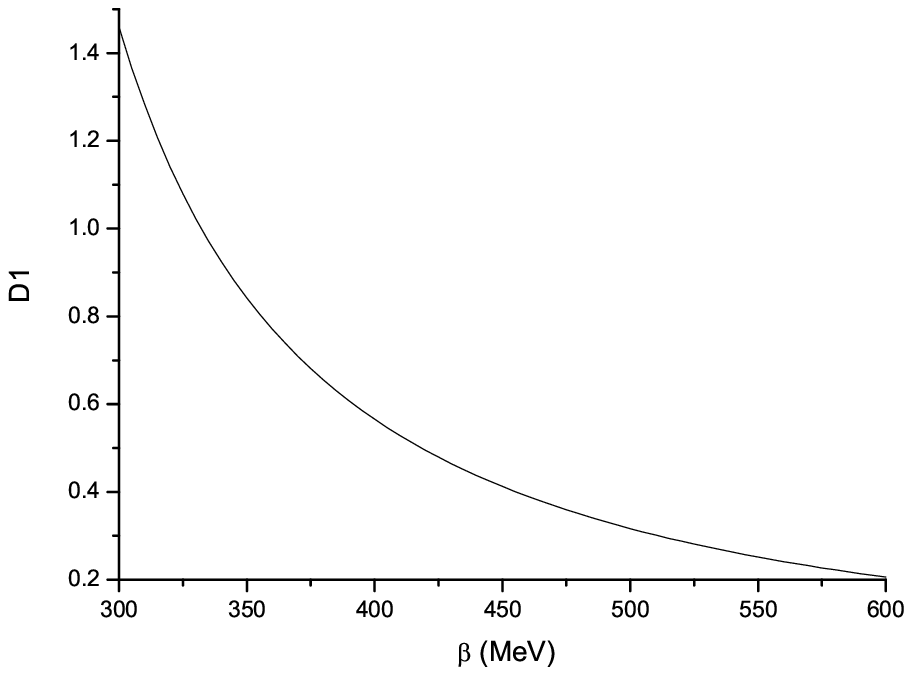}
 \includegraphics[scale=0.5]{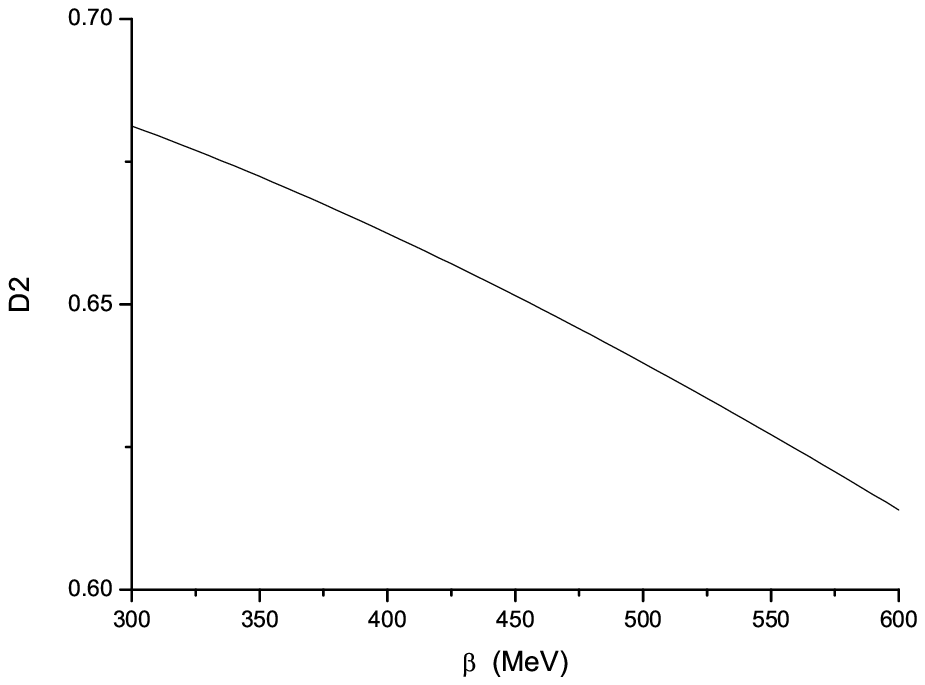}
 \figcaption{\label{Figure-1}
    $D/S$ ratios for $D_1$ and $D_2$.}
\end{center}

With respect to $D_1$ ratio, it decreases with an increasing $\beta$. When
$\beta$ rises up to $524$MeV, the ratio regenerates the
experimental value $0.277$. Nonetheless, as to $D_2$ ratio, the
numerical value keeps its opposite sign since this ratio changes
rather slowly with $\beta$.

Based on the fact of the dominance of the scalar interaction, a
scalar-kernel-scalar ($sKs$) decay model was proposed \cite{L7}. As a
result we consider only the contribution from scalar interaction while
leaving the vector interaction aside. The best-fitted value for $M_g$
now becomes $597$MeV and the fitted decay widths are listed in the
$\Gamma_3$ column in Table~\ref{Table-2}.

One of the advantages of considering scalar interaction alone is that
the $D/S$ ratios are greatly improved. Now these ratios turn to be:
\begin{align}
  D_1=&-\frac{5\sqrt{2}p^2_f}{5p^2_f-24\beta^2} .\\
  D_2=&\frac{5p^2_f}{\sqrt{2}(5p^2_f-24\beta^2)}.
\end{align}

We obtain $D_1=0.264$ and $D_2=-0.140$ when $\beta=400$MeV, well fit
to the experimental results.  Another improvement lies in the specific
channel, ${}^3P_0\rightarrow{}^1S_0+{}^1S_0$. As in the process
$f_0(1370)\rightarrow \pi\pi$, with the negligence of vector
interaction, the decay width becomes a reasonable value $318.5$MeV.

\section{Summary}

To summarize, we have studied a decay model based on the potential
quark model. The model incorporates the decay interactions of scalar
and vector quark currents which are in accordance with the confining and
OGE potentials in the quark model. In the non-relativistic limit, the
massive gluon propagator is assumed and the decay interactions are
reduced to four fermion interactions. In this framework, we have
calculated 34 decay channels. The results fit the experimental data
comparable to the popular ${}^3P_0$ decay model if the $SU(3)$ flavor
symmetry is assumed in the decay processes.  Meanwhile the results
also show the dominance of the scalar interaction in most of the decay
channels.  Besides, the scalar interaction is also preferred by the
$D/S$ ratios of $b_1\rightarrow\omega +\pi$ and $a_1\rightarrow
\rho+\pi$.  Thus we have calculated decay widths with only scalar
interaction (the $sKs$ model which is quite similar to the ${}^3P_0$
model since the $q\bar q$ scalar current produces a ${}^3P_0$ quark
pair). It seems that scalar interaction alone is able to give a crude
estimation of most decay widths.

\acknowledgments{
  We would like to thank Prof. Shi-Lin Zhu for useful discussions.
}

\end{multicols}
\vspace{10mm}

\appendix
\renewcommand{\theequation}{A\arabic{equation}}
\setcounter{equation}{0}
\begin{multicols}{2}
\section*{Appendix}
\begin{small}

\section{Amplitudes and Overlapping Integrals for Some Channels}

The partial wave decay amplitude $A\to B+C$ can be expressed as
\renewcommand{\theequation}{A\arabic{equation}}
\begin{equation}
  \mathcal{M}^{LS}=\frac{\sqrt{8E_AE_BE_C}}{24\pi^5} C_f M^{LS}.
\end{equation}

where $C_f$ is the flavor factor:

\begin{equation}
  C_f^2 = (2T_B+1)(2T_C+1)\begin{Bmatrix}
    T_A & T_B & T_C \\
    t & t_2 & t_1
  \end{Bmatrix}^2.
\end{equation}

where $T_A$, $T_B$, $T_C$ are the iso-spins of mesons $A$, $B$, $C$,
respectively.  $t_1$, $t_2$, $t$ are the iso-spins of quarks labeled
as $1$, $2$, $3$ in Figure~\ref{Fig-1}, respectively.  Similarly, in
the following, the masses of quarks $1$, $2$, $3$ will be denoted by
$m_1$, $m_2$, $m$, respectively.

The decay amplitudes $M^{LS}$ can be split into two parts which include scalar
$M^{LS}_s$ and vector $M^{LS}_v$.  Furthermore, each $M^{LS}_i$ would part
into four components:
\begin{equation}
  M^{LS}_i=M^{LS}_i(a)+M^{LS}_i(b)+M^{LS}_i(c)+M^{LS}_i(d).
\end{equation}
according to  Figure~\ref{Fig-1}. In the following, we will only present
the formulae for $M^{LS}_i(a)$. $M^{LS}_i(b)$ is related to $M^{LS}_i(a)$ by a
charge conjugate:
\begin{eqnarray}
  M^{LS}_i(b;A\to B+C) &=& (-1)^{J_B+J_C-S+S_A+S_B+S_C+1}
   \nonumber\\[1mm]&&\times M^{LS}_i(a;\bar{A}\to \bar{C}+\bar{B}).
\end{eqnarray}
where $\bar{A}$, $\bar{B}$, $\bar{C}$ are the charge conjugates of
$A$, $B$, $C$, respectively. The $M^{LS}_i(c)$, $M^{LS}_i(d)$ are
related to $M^{LS}_i(a)$, $M^{LS}_i(b)$ by the exchange of final particles
$B$ and $C$.

Since the decay interaction is four-fermion interaction, the spatial
overlap integrals involve:

\begin{eqnarray}
  p_{AC}(m_A,m_C)&=&\int d\bm{k}\psi^{\ast}_{n_Cl_Cm_C}(\bm{k})\nonumber\\&&\times
  \psi_{n_Al_Am_A}(\bm{k}+\xi p_f \hat{\bm{z}}).\nonumber\\
  v_{AC}(m_A,m_C,m)&=&\int d\bm{k}\psi^{\ast}_{n_Cl_Cm_C}(\bm{k})\nonumber\\&&\times
  \psi_{n_Al_Am_A}(\bm{k}+\xi p_f \hat{\bm{z}})k_m\nonumber.\\
  p_B(m_B)&=&\int d\bm{k}\psi^{\ast}_{n_Bl_Bm_B}(\bm{k}).\nonumber\\
  v_B(m_B,m)&=&\int d\bm{k}\psi^{\ast}_{n_Bl_Bm_B}(\bm{k})k_m.  \label {eq9}
\end{eqnarray}

where $\xi=\frac{m_2}{m_2+m}$, and
\begin{equation}
  k_m=
  \begin{cases}
    -\frac{1}{\sqrt{2}}(k_x+ik_y)
    &m=1\\
    k_z
    &m=0\\
    +\frac{1}{\sqrt{2}}(k_x-ik_y)
    &m=-1
  \end{cases}
\end{equation}
All the spatial wave functions $\psi_{nlm}$ are taken to be the simple
harmonic oscillator (SHO) wave functions with the oscillator parameters
$\beta_A$, $\beta_B$, $\beta_C$ for mesons $A$, $B$, $C$
respectively. We have
\begin{align}
  p_B(m_B) =& \frac{\delta(l_B,0)}{n_B!}
  (4\pi\beta^2_{B})^{\frac{3}{4}} .\\
  v_B(m_B,m) =& \frac{\delta(l_B,1)\delta(m_B,m)}{n_B!}
  4\pi^{\frac34}\beta_{B}^{\frac{5}{2}}.
\end{align}
Let
\begin{equation}
  \eta\equiv
  \left(\frac{2\beta_{A}\beta_{C}}{\beta^2_{A}
    +\beta^2_{C}}\right)^{\frac{3}{2}}e^{-\frac{\xi^2p^2_{f}}{2(\beta^2_{A}+\beta^2_{C})}}.
\end{equation}

Below we list the non-vanishing integrals $p_{AC}$ and $v_{AC}$ relevant
to our work.
\begin{itemize}
\item $1S\rightarrow1S$
  \begin{eqnarray*}
    p_{AC}(0,0)&=&\eta\\
    v_{AC}(0,0,0)&=&-\frac{\xi\eta\beta^2_{C}}{\beta^2_{A}+\beta^2_{C}}p_f
  \end{eqnarray*}
\item $2S\rightarrow1S$
  \begin{eqnarray*}
    p_{AC}(0,0)&=&\frac{(3\beta_A^4-3\beta_C^4-2p_f^2\beta_A^2\xi^2)\eta}
    {\sqrt{6}(\beta^2_{A}+\beta^2_{C})^2}\\
    v_{AC}(0,0,0)&=& \frac{p_f\beta^2_C\xi\eta}{\sqrt{6}(\beta^2_{A}+\beta^2_{C})^3}(-7\beta^4_A
      +3\beta^4_C\\[1mm]&&-4\beta^2_A\beta^2_C+2p^2_f\xi^2\beta^2_A)
  \end{eqnarray*}
\item $1P\rightarrow1S$
  \begin{eqnarray*}
    p_{AC}(0,0)&=&\frac{\sqrt{2}\eta\xi\beta_A p_{f}}{\beta^2_A+\beta^2_C}\\
    v_{AC}(0,0,0)&=&\frac{\sqrt{2}\eta\beta_A\beta^2_C(\beta^2_A
      +\beta^2_C-p^2_f\xi^2)}{(\beta^2_A+\beta^2_C)^2}\\
    v_{AC}(1,0,-1)&=&v_{AC}(-1,0,1)=\frac{-\sqrt{2}\eta\beta_A\beta^2_C}
    {(\beta^2_A+\beta^2_C)}
  \end{eqnarray*}
\item $1D\rightarrow1S$
  \begin{eqnarray*}
    p_{AC}(0,0)&=&\frac{2p^2_f\beta^2_A\xi^2\eta}
    {\sqrt{3}(\beta^2_A+\beta^2_C)^2}\\
    v_{AC}(0,0,0)&=&\frac{2p_f\beta^2_A\beta^2_C\xi\eta(2\beta^2_A
      +2\beta^2_C-p^2_f\xi^2)}{\sqrt{3}(\beta^2_A+\beta^2_C)^3}\\
    v_{AC}(1,0,-1)&=&v_{AC}(-1,0,1)=\frac{-2p_f\beta^2_A\beta^2_C\eta\xi}
    {(\beta^2_A+\beta^2_C)^2}
  \end{eqnarray*}
  \end{itemize}
\end{small}
\end{multicols}
\ruleup
\begin{itemize}
\item $1D\rightarrow1P$
  \begin{eqnarray*}
    p_{AC}(0,0)&=&\frac{2\sqrt{2}p_f\xi\eta\beta^2_A\beta_C(2\beta^2_A
      +2\beta^2_C-p^2_f\xi^2)}{\sqrt{3}(\beta^2_A+\beta^2_C)^3}\\
    p_{AC}(1,1)&=&p_{AC}(-1,-1)=\frac{2\sqrt{2}p_f\eta\xi\beta^2_A\beta_C}
    {(\beta^2_A+\beta^2_C)^2}\\
    v_{AC}(0,0,0)&=&\frac{2\sqrt{2}\beta^2_A\beta_C\eta}
    {\sqrt{3}(\beta^2_A+\beta^2_C)^4}[2\beta^6_C-4p^2_f\beta_C^4\xi^2
    +p^4_f\beta^2_C\xi^4+\beta^4_A(2\beta^2_C+p^2_f\xi^2)
    +\beta^2_A(4\beta^4_C-3p^2_f\beta^2_C\xi^2)]\\
    v_{AC}(2,1,-1)&=&v_{AC}(-2,-1,1)=-\frac{4\beta^2_A\beta^3_C\eta}
    {(\beta^2_A+\beta^2_C)^2}\\
    v_{AC}(1,0,-1)&=&v_{AC}(-1,0,1)=-\frac{2\sqrt{2}\beta^2_A\beta^3_C
      \eta(\beta^2_A+\beta^2_C-\xi^2p^2_f)}{(\beta^2_A+\beta^2_C)^3}\\
    v_{AC}(1,1,0)&=&v_{AC}(-1,-1,0)=-v_{AC}(1,0,-1)\\
    v_{AC}(0,1,1)&=&v_{AC}(0,-1,-1)=-\frac{2\sqrt{2}\eta\beta^2_A
      \beta_C[\beta^4_C+\beta^2_A(\beta^2_C-\xi^2p^2_f)]}{\sqrt{3}(\beta^2_A
      +\beta^2_C)^3}
  \end{eqnarray*}
\item $1F\rightarrow1S$
  \begin{eqnarray*}
    p_{AC}(0,0)&=&\frac{2\sqrt{2}\eta\xi^3\beta^3_Ap^3_f}
    {\sqrt{15}(\beta^2_A+\beta^2_C)^3}\\
    v_{AC}(0,0,0)&=&\frac{2\sqrt{2}\eta p^2_f\beta^3_A\beta^2_C\xi^2(3\beta^2_A
      +3\beta^2_C-p^2_f\xi^2)}{\sqrt{15}(\beta^2_A+\beta^2_C)^4}\\
    v_{AC}(1,0,-1)&=&v_{AC}(-1,0,1)=\frac{-4\eta p^2_f\beta^3_A\beta^2_C\xi^2}
    {\sqrt{5}(\beta^2_A+\beta^2_C)^3}
  \end{eqnarray*}
\end{itemize}

We further introduce some useful combinations:

\begin{align}
  A_L =& p_f p_B(0) \sum_{m_Am_C} p_{AC}(m_A,m_C)
  \braket{l_C m_C L 0 | l_A m_A} \\
  B_{LJ} =& \sqrt{2l_A+1} p_B(0) \sum_{m_Am_Cm} v_{AC}(m_A,m_C,m)
  \braket{l_C m_C L 0 | J m_C} \braket{l_A m_A 1 m | J m_C}
\end{align}

The relevant partial wave amplitudes are given in subsections.

\subsection{$S \rightarrow S+S$}

\begin{itemize}
\item ${}^3S_1\rightarrow {}^1S_0+{}^1S_0$
  \begin{eqnarray*}
    M^{10}_s(a)&=&-\frac{\sqrt \pi b}{\sqrt 2 M^4_g}\left[(\frac{1}{m+m_1}
    +\frac{1}{m+m_2})A_0-\frac1m B_{11}\right]\\
    M^{10}_{v}(a)&=&-\frac{\sqrt{2\pi}\alpha_s}{3M^2_g}\left[\frac{m_1(m_1+m_2)
      -m(m_1-3m_2)}{m_1(m+m_1)(m+m_2)}A_0+(\frac{3}{m_1}-\frac1m)B_{11}\right]
  \end{eqnarray*}

\item ${}^3S_1\rightarrow {}^3S_1+{}^1S_0$
  \begin{eqnarray*}
    M^{11}_s(a)&=&\sqrt2 M_s^{10}(a;{}^3S_1\rightarrow {}^1S_0+{}^1S_0)\\
    M^{11}_{v}(a)&=&-\frac{2\sqrt\pi\alpha_s}{3M^2_g}
    \left[(\frac{m_1+m_2}{m_1(m+m_2)})A_0
    +(\frac1{m_1}-\frac1m)B_{11}\right]
  \end{eqnarray*}

\item ${}^3S_1\rightarrow {}^1S_0+{}^3S_1$
  \begin{eqnarray*}
    M^{11}_s(a)&=&-\sqrt2 M_s^{10}(a;{}^3S_1\rightarrow {}^1S_0+{}^1S_0)\\
    M^{11}_{v}(a)&=&-\sqrt2 M_v^{10}(a;{}^3S_1\rightarrow {}^1S_0+{}^1S_0)
  \end{eqnarray*}

\item ${}^1S_0\rightarrow {}^3S_1+{}^1S_0$
  \begin{eqnarray*}
    M^{11}_s(a)&=&-\sqrt{3}M_s^{10}(a;{}^3S_1\rightarrow {}^1S_0+{}^1S_0)\\
    M^{11}_{v}(a)&=&\frac{\sqrt{2\pi}\alpha_s}{\sqrt3M^2_g}\left[\frac{m(3m_1-m_2)
      +m_1(m_1+m_2)}{m_1(m+m_1)(m+m_2)}A_0-(\frac1m+\frac{1}{m_1})B_{11}\right]
  \end{eqnarray*}

\item ${}^1S_0\rightarrow {}^1S_0+{}^3S_1$
  \begin{eqnarray*}
    M^{11}_s(a)&=&-\sqrt{3}M_s^{10}(a;{}^3S_1\rightarrow {}^1S_0+{}^1S_0)\\
    M^{11}_{v}(a)&=&-\sqrt3 M_v^{10}(a;{}^3S_1\rightarrow {}^1S_0+{}^1S_0)
  \end{eqnarray*}
\end{itemize}

\subsection{$P \rightarrow S+S$}

\begin{itemize}
\item ${}^3P_0\rightarrow {}^1S_0+{}^1S_0$
  \begin{eqnarray*}
    M^{00}_s(a)&=&\frac{\sqrt\pi b}{\sqrt2 M^4_g}\left[(\frac{1}{m+m_1}
    +\frac{1}{m+m_2})A_1+\frac{1}m B_{00}\right]\\
    M^{00}_{v}(a)&=&\frac{\sqrt{2\pi}\alpha_s}{3M^2_g}\left[\frac{m_1(m_1-m)
      +m_2(m_1+3m)}{m_1(m+m_1)(m+m_2)}A_1-(\frac{3}{m_1}-\frac1m)
    B_{00}\right]
  \end{eqnarray*}

\item ${}^3P_2\rightarrow {}^1S_0+{}^1S_0$
  \begin{eqnarray*}
    M^{20}_s(a)&=&-\frac{\sqrt{\pi} b}{\sqrt5 M^4_g}\left[(\frac{1}{m+m_1}
      +\frac{1}{m+m_2})A_1-\frac{\sqrt2}{2m}B_{22}\right]\\
    M^{20}_{v}(a)&=&-\frac{\sqrt{\pi}\alpha_s}{3\sqrt5M^2_g}
    \left[\frac{2m_1(m_1-m)+2m_2(m_1+3m)}{m_1(m+m_1)(m+m_2)}A_1
      +\sqrt2(\frac{3}{m_1}-\frac1m)B_{22}\right]
  \end{eqnarray*}

\item ${}^3P_2\rightarrow {}^3S_1+{}^1S_0$
  \begin{eqnarray*}
    M^{21}_s(a)&=&\frac{\sqrt6}{2}M^{20}_s(a;{}^3P_2\rightarrow {}^1S_0+{}^1S_0)\\
    M^{21}_{v}(a)&=&-\frac{\sqrt\pi\alpha_s}{\sqrt{30}M^2_g}
    \left[\frac{2(m_1+m_2)}{m_1(m+m_2)}A_1+\sqrt2(\frac{1}{m_1}-\frac1m)
      B_{22}\right]
  \end{eqnarray*}

\item ${}^3P_2\rightarrow {}^1S_0+{}^3S_1$
  \begin{eqnarray*}
    M^{21}_s(a)&=&-\frac{\sqrt6}{2}M^{20}_s(a;{}^3P_2\rightarrow {}^1S_0+{}^1S_0)\\
    M^{21}_{v}(a)&=&\frac{\sqrt\pi\alpha_s}{\sqrt{30}M^2_g}
    \left[\frac{2m_1(m_1-m)+2m_2(m_1+3m)}{m_1(m+m_1)(m+m_2)}A_1
      +\sqrt2(\frac{3}{m_1}-\frac1m)B_{22}\right]
\end{eqnarray*}
\end{itemize}

\subsection{$D \rightarrow S+S$}

\begin{itemize}
\item ${}^1D_2\rightarrow {}^3S_1+{}^3S_1$
  \begin{eqnarray*}
    M^{11}_s(a)&=&-\frac{\sqrt{3\pi}b}{5\sqrt{2}M^4_g}\left[2(\frac{1}{m+m_1}
      +\frac{1}{m+m_2})A_2+\frac{\sqrt{2}}mB_{11}\right]\\
    M^{11}_{v}(a)&=&-\frac{\sqrt{2\pi}\alpha_s}{5\sqrt{3}M^2_g}
    \left[\frac{2(m_1+m_2)}{m_1(m+m_2)}A_2
      +\sqrt2(\frac1m-\frac{1}{m_1})B_{11}\right]\\
    M^{31}_s(a)&=&\frac{\sqrt{\pi}b}{5M^4_g}\left[3(\frac{1}{m+m_1}
      +\frac{1}{m+m_2})A_2-\frac{\sqrt3}mB_{33}\right]\\
    M^{31}_{v}(a)&=&\frac{2\sqrt{\pi}\alpha_s}{5M^2_g}\left[\frac{m_1+m_2}
      {m_1(m+m_2)}A_2+\frac1{\sqrt3}(\frac{1}{m_1}-\frac1m)B_{33}\right]
  \end{eqnarray*}

\item ${}^1D_2\rightarrow {}^3S_1+{}^1S_0$
  \begin{eqnarray*}
    M^{11}_s(a)&=&\frac{1}{\sqrt2}M^{11}_s(a;{}^1D_2\rightarrow {}^3S_1+{}^3S_1)\\
    M^{11}_{v}(a)&=&-\frac{\sqrt{\pi}\alpha_s}{5\sqrt{3}m^2_g}
    \left[\frac{2m(3m_1-m_2)+2m_1(m_1+m_2)}{m_1(m+m_1)(m+m_2)}A_2
      +\sqrt2(\frac1m+\frac{1}{m_1})B_{11}\right]\\
    M^{31}_s(a)&=&\frac{1}{\sqrt2}M^{31}_s(a;{}^1D_2\rightarrow {}^3S_1+{}^3S_1)\\
    M^{31}_{v}(a)&=&\frac{\sqrt{2\pi}\alpha_s}{5m^2_g}
    \left[\frac{m(3m_1-m_2)+m_1(m_1+m_2)}{m_1(m+m_1)(m+m_2)}A_2
      -\frac1{\sqrt3}(\frac1m+\frac{1}{m_1})B_{33}\right]
  \end{eqnarray*}

\item ${}^1D_2\rightarrow {}^1S_0+{}^3S_1$
  \begin{eqnarray*}
    M^{11}_s(a)&=&\frac{1}{\sqrt2}M^{11}_s(a;{}^1D_2\rightarrow {}^3S_1+{}^3S_1)\\
    M^{11}_{v}(a)&=&\frac{\sqrt{\pi}\alpha_s}{5\sqrt{3}m^2_g}
    \left[\frac{2m(m_1-3m_2)-2m_1(m_1+m_2)}{m_1(m+m_1)(m+m_2)}A_2
      +\sqrt2(\frac{3}{m_1}-\frac1m)B_{11}\right]\\
    M^{31}_s(a)&=&\frac{1}{\sqrt2}M^{31}_s(a;{}^1D_2\rightarrow {}^3S_1+{}^3S_1)\\
    M^{31}_{v}(a)&=&\frac{\sqrt{2\pi}\alpha_s}{5m^2_g}\left[\frac{m_1(m_1+m_2)
        -m(m_1-3m_2)}{m_1(m+m_1)(m+m_2)}A_2
      +\sqrt3(\frac{1}{m_1}-\frac{1}{3m})B_{33}\right]
  \end{eqnarray*}

\item ${}^3D_3\rightarrow {}^1S_0+{}^1S_0$
  \begin{eqnarray*}
    M^{30}_s(a)&=&-\frac{5}{\sqrt{70}}
    M^{31}_s(a;{}^1D_2\rightarrow {}^3S_1+{}^3S_1)\\
    M^{30}_{v}(a)&=&\frac{\sqrt{2\pi}\alpha_s}{\sqrt{35}M^2_g}
    \left[\frac{m(m_1-3m_2)-m_1(m_1+m_2)}{m_1(m+m_1)(m+m_2)}A_2
      -\sqrt3(\frac{1}{m_1}-\frac{1}{3m})B_{33}\right]
  \end{eqnarray*}

\item ${}^3D_3\rightarrow {}^3S_1+{}^1S_0$
  \begin{eqnarray*}
    M^{31}_s(a)&=&-\sqrt{\frac{10}{21}}
    M^{31}_s(a;{}^1D_2\rightarrow {}^3S_1+{}^3S_1)\\
    M^{31}_{v}(a)&=&-2\frac{\sqrt{2\pi}\alpha_s}{\sqrt{105}M^2_g}
    \left[\frac{m_1+m_2}{m_1(m+m_2)}A_2
      +\frac1{\sqrt3}(\frac{1}{m_1}-\frac1m)B_{33}\right]
  \end{eqnarray*}

\item ${}^3D_3\rightarrow {}^1S_0+{}^3S_1$
  \begin{eqnarray*}
    M^{31}_s(a)&=&\sqrt{\frac{10}{21}}
    M^{31}_s(a;{}^1D_2\rightarrow {}^3S_1+{}^3S_1)\\
    M^{31}_{v}(a)&=&2\frac{\sqrt{2\pi}\alpha_s}{\sqrt{105}M^2_g}
    \left[\frac{m_1(m_1-m)+m_2(m_1+3m)}{m_1(m+m_1)(m+m_2)}A_2
      +\sqrt3(\frac{1}{m_1}-\frac{1}{3m})B_{33}\right]
  \end{eqnarray*}
\end{itemize}

\subsection{$D \rightarrow P+S$}

\begin{itemize}
\item ${}^1D_2\rightarrow {}^3P_2+{}^1S_0$
  \begin{equation*}
    M^{LS}_s(a)=M^{LS}_{v}(a)=0
  \end{equation*}

\item ${}^1D_2\rightarrow {}^1S_0+{}^3P_2$
  \begin{eqnarray*}
    M^{02}_s&=&-\frac{\sqrt\pi b}{5\sqrt2mM^4_g}
    \left[\sqrt3\frac{m(m_1+m_2+2m)}{(m+m_1)(m+m_2)}
      A_1+B_{01}\right]\\
    M^{02}_{v}&=&-\frac{\sqrt{2\pi} \alpha_s}{15M^2_g}\left[
      \frac{m_1(m_1-m)+m_2(m_1+3m)}
      {m_1(m+m_1)(m+m_2)}\sqrt{3}A_1+(\frac1m-\frac{3}{\sqrt2m_1})B_{01}\right]\\
    M^{22}_s&=&\frac{\sqrt\pi b}{2\sqrt{35}mM^4_g}\left[
      \frac{\sqrt2m(m_1+m_2+2m)}{(m+m_1)(m+m_2)}A'-B'\right]\\
    M^{22}_{v}&=&\frac{\sqrt\pi \alpha_s}{3\sqrt{35}M^2_g}\left[
      \frac{\sqrt 2[m_1(m_1-m)+m_2(m_1+3m)]}{m_1(m+m_1)(m+m_2)}A'
      +(\frac{3}{m_1}-\frac1m)B' \right]\\
    M^{42}_s&=&\frac{\sqrt{6\pi} b}{10\sqrt{7}mM^4_g}\left[
      \frac{2\sqrt{7}m(m_1+m_2+2m)}
      {(m+m_1)(m+m_2)}A_3-3B_{43}\right]\\
    M^{42}_{v}&=&\frac{\sqrt{6\pi} \alpha_s}{15\sqrt{7}M^2_g}
    \left[\frac{2\sqrt7[m_1(m_1-m)+m_2(m_1+3m)]}{m_1(m+m_1)(m+m_2)} A_3
      +3(\frac{3}{m_1}-\frac1m)B_{43}\right]
  \end{eqnarray*}
  where
  \begin{align*}
    A'=& p_f p_B(0)[\sqrt3 p_{AC}(1,1)+2p_{AC}(0,0)] \\
    B'=& p_B(0)[4\sqrt3 v_{AC}(2,1,-1)
    +\sqrt6 v_{AC}(1,1,0)
    -\sqrt6 v_{AC}(1,0,-1)
    -2\sqrt2 v_{AC}(0,1,1)
    +2\sqrt2 v_{AC}(0,0,0)]
  \end{align*}
\end{itemize}

\subsection {$F \rightarrow S+S$}

\begin{itemize}
\item ${}^3F_4\rightarrow {}^1S_0+{}^1S_0$
  \begin{eqnarray*}
    M^{40}_s&=&\frac{\sqrt{2\pi}b}{\sqrt{21}M^4_g}\left[
      -(\frac{1}{m+m_1}+\frac{1}{m+m_2})A_3+\frac{1}{2m}B_{44}\right]\\
    M^{40}_{v}&=&\frac{\sqrt{2\pi}\alpha_s}{3\sqrt{21}M^2_g}\left[
      \frac{2m(m_1-3m_2)-2m_1(m_1+m_2)}{m_1(m+m_1)(m+m_2)}A_3
      -3(\frac{1}{m_1}-\frac{1}{3m})B_{44}\right]\\
  \end{eqnarray*}
\end{itemize}
\vspace{-1mm}
\centerline{\rule{80mm}{0.1pt}}
\vspace{2mm}
\begin{multicols}{2}
\bibliography{decay}
\end{multicols}
\end{document}